\documentclass{article}
\usepackage{amsfonts}
\usepackage{epsfig}
\begin{document}

\title{\textsc{Markov Chain Methods For Analyzing Complex Transport Networks}}

\vspace{1cm}

\author{ D. Volchenkov and Ph. Blanchard
\vspace{0.5cm}\\
{\it  BiBoS, University Bielefeld, Postfach 100131,}\\
{\it D-33501, Bielefeld, Germany} \\
{\it Phone: +49 (0)521 / 106-2972 } \\
{\it Fax: +49 (0)521 / 106-6455 } \\
{\it E-Mail: VOLCHENK@Physik.Uni-Bielefeld.DE}}

\date{\today}
\maketitle

\begin{abstract}
We have developed a steady state theory of complex
transport networks used to model the flow of commodity,
 information, viruses, opinions, or traffic. Our
 approach is based on the use of the Markov chains
 defined on the graph representations of transport
  networks allowing for the effective network design,
   network performance evaluation, embedding,
    partitioning, and network fault tolerance
     analysis. Random walks embed graphs into
     Euclidean space in which distances and
     angles acquire a clear statistical
      interpretation. Being defined on the dual graph
      representations of transport networks
       random walks describe the equilibrium
       configurations of not random commodity
       flows on primary graphs. This theory
       unifies many network concepts into one
       framework and can also be elegantly
       extended to describe networks represented
       by directed graphs and multiple interacting networks.
\end{abstract}

\vspace{0.5cm}

\vspace{0.5cm}

\leftline{\textbf{ PACS codes: 89.75.Fb, 89.75.-k, 89.90.+n} }
 \vspace{0.5cm}

\leftline{\textbf{ Keywords:} Random walks; complex networks;  traffic equilibrium. }

\section{Introduction}
\label{sec:Introduction}
 \noindent

Transport networks are used
to model the flow of commodity,
 information, viruses, opinions, or traffic.
They typically represent the
 networks of roads, streets, pipes,
aqueducts, power lines, or
 nearly any structure
 which permits either vehicular
movement or flow of some commodity,
  products, goods or service.
The major aim of the analysis is to
determine the structure and properties
 of
 transport networks that are important
for  the emergence of
 complex flow patterns
of vehicles
(or people) through the network
such as the {\it  Braess paradox} \cite{Braess}.
This counter-intuitive
phenomenon occurs when
adding more resources to a transportation network
(say, a new road or a bridge)
deteriorates the quality of traffic by creating
worse delays for the drivers,
rather than alleviate it. The Braess paradox
has been
observed in the street vehicular
traffic of New York City and Stuttgart, \cite{Kolata}.

The paper is partitioned into three sections. In
Sec.~\ref{sec:Euclidean}, we demonstrate that Markov chains arise
naturally in the problem of network equilibriums. Random walks
embed connected undirected graphs into Euclidean space that can be
used in order to investigate them. In
Sec.~\ref{sec:Thermodynamics}, we discuss the thermodynamics of
random walks defined on the undirected graph representations of
transport networks. In Sec.~\ref{sec:Directed}, we extend the
approach to networks represented by directed graphs and to
multiple interacting networks. We conclude in the last section.

\section{Euclidean Space Associated to Transport Networks}
\label{sec:Euclidean}
 \noindent

The main goal of
complex network theory
is to
study relationships between
 parts of complex systems that gives
rise to the patterns of collective
behavior and of
 interactions between complex systems with
their environments.

\subsection{Traffic equilibrium, space syntax, and random walks}
\label{subsec:equilibrium}
 \noindent

Given a connected undirected graph $G(V,E)$, in which $V$ is the
set of nodes and $E$ the set of edges, we introduce the traffic
 volume $f: E\to(0,\infty[$ through every edge $e\in E$. It then follows
  from the Perron-Frobenius theorem that the linear equation
\begin{equation}
\label{Lim_equilibrium}
f(e)\,=\, \sum_{e'\,\in\, E}\,f(e')\,\exp\left(\,-h\,\ell\left(e'\right)\,\right)
\end{equation}
has a unique positive solution $f(e)>0$, for every edge $e\in E$, for a
fixed positive constant $h>0$ and a chosen set of positive  {\it metric
 length} distances $\ell(e)>0$. This solution is naturally identified
 with the traffic equilibrium state of the transport network defined on
  $G$, in which the permeability of edges depends upon their lengths.
The parameter $h$  is called the volume entropy of the graph $G$, while
 the volume of $G$ is defined as the sum
\[
\mathrm{Vol}(G)\,=\,\frac 12\,\sum_{e\in E}\,\ell(e).
\]
The volume entropy $h$ is defined to be the exponential growth of the balls in
a universal covering tree for $G$ with the lifted metric, \cite{Manning}-\cite{Lim:2005}.

The degree of a node $v\in V$ is the number of its neighbors in
$G$, $\deg(v)=k_v$. It has been shown in \cite{Lim:2005} that
among all undirected connected graphs of normalized volume,
$\mathrm{Vol}(G)=1$, which are not cycles and $k_v\ne 1$ for all
nodes,
 the minimal possible value of the volume entropy,
$\min(h)=\frac 12\sum_{v\in V}k_v\,\log\left(k_v-1\right)$  is attained
for the length distances
\begin{equation}
\label{ell_min}
\ell(e)\,=\,\frac {\log\left(\left(k_{i(e)}-1\right)
\left(k_{t(e)}-1\right)\right)}{2\,\min(h)},
\end{equation}
where $i(e)\in V$ and $t(e)\in V$ are the initial and terminal vertices
of the edge $e\in E$ respectively. It is then obvious that
substituting
(\ref{ell_min})
and $\min(h)$ into (\ref{Lim_equilibrium}) the
operator $\exp\left(-h \ell(e')\right)$ is given by
 a symmetric Markov transition operator,
\begin{equation}
\label{Markov_transition}
f(e)\,=\, \sum_{e'\,\in\, E}\,\frac{f(e')}{\sqrt{\left(k_{i(e')}-1\right)
\left(k_{t(e')}-1\right)}},
\end{equation}
which rather describes time reversible random walks over edges than over
nodes.
In other words, we are invited to consider random walks on the dual graphs.
 The flows satisfying (\ref{Lim_equilibrium}) with the operator
(\ref{Markov_transition}) meet the mass conservation property,
$\sum_{i\sim j}f_{ij}=\pi_j$, $\sum_{j\in V}\pi_j=1$.
The Eq.(\ref{Markov_transition}) unveils the indispensable role Markov's
 chains defined on edges play in equilibrium traffic modelling and exposes
 the degrees of nodes as a key determinant of the transport networks properties.

The notion of traffic equilibrium had been introduced by J.G. Wardrop in
 \cite{Wardrop:1952} and then generalized in \cite{Beckmann:1956} to a
 fundamental concept of   network equilibrium. Wardrop's traffic
 equilibrium  is strongly tied to the human apprehension of space since
  it is required that all travellers have enough knowledge of the
  transport network they use. The human perception of places is not
  an entirely Euclidean one, but are rather related to the perceiving
  of the vista spaces (streets and squares) as single units and of the
   understanding of the topological relationships between these vista spaces,
   \cite{Kuipers}. Decomposition of city space into a complete set of
   intersecting vista spaces produces a spatial network which we call
   the {\it dual} graph representation of a city. Therein, the relations
    between streets treated as nodes are traced through their  junctions
     considered as edges.

Dual city graphs are extensively investigated within the concept of
{\it space syntax}, a theory developed in the late 1970s, that seeks to
reveal the mutual effects of complex spatial urban networks on society
and vice versa, \cite{Hillier:1984,Hillier:1999}.
Spatial perception
 that shapes peoples understanding of how
a place is organized  determines eventually the pattern of local movement
 which is predicted
by the space syntax method with surprising accuracy \cite{Penn:2001}.

\subsection{Euclidean space of undirected graphs associated to random walks }
\label{subsec:Euclidean_space}
 \noindent

Any graph representation  naturally arises as the outcome of a categorization,
when we abstract a real world system by eliminating all but one of its features
 and by  grouping together things (or places) sharing a common attribute.
All elements called
 nodes that fall into one and the same group $V$ are considered as essentially
 identical; permutations of them within the  group are of no consequence.
The symmetric group $\mathbb{S}_{N}$ consisting of all permutations of $N$
elements
($N$ being the cardinality of the set $V$) constitute the symmetry group of $V$.
If we denote by $E\subseteq V\times V$ the set of ordered pairs of nodes called
edges, then  a graph is a map $G(V,E): E \to K\subseteq\mathbb{R}_{\,+}$
(we suppose that the graph has no multiple edges).

The nodes of $G(V,E)$ can be weighted with respect to some  {\it  measure}
 $m=\sum_{i\in V} m_i \delta_i,$ specified by a set of positive numbers $m_i> 0$.
  The space $\ell^2(m)$ of square-assumable functions with respect to the
   measure $m$ is a {\it Hilbert space} $\mathcal{H}$.
Among all linear operators defined on $\mathcal{H}$,
   those  {\it invariant} under the permutations of nodes are particularly
   interesting since they reflect the symmetry of the graph. Although there are
   infinitely many such operators, only those which maintain  conservation
   of a quantity may describe a physical process. The Markov transition
   operators which share the property of  {\it probability conservation}
   considered in the theory of random walks on graphs are among them.
Another example is given by the Laplace operators  satisfying the
{\it mean value}
 property ({\it mass conservation}) \cite{Smola2003}.

 Markov's operators on Hilbert space  form the natural language of complex networks theory.
Being defined on connected undirected graphs, a Markov transition operator $T$ has a unique
 equilibrium state $\pi$ ( stationary distribution of the random walk)
 such that $\pi T=\pi$ and $\pi=\lim_{t\to\infty}\,\sigma\, T^{\,t}$ for
  any density $\sigma\in \mathcal{H}$ ($\sigma_i\geq 0$,
$\sum_{i\in V} \sigma_i=1$). There is a unique measure
$m_\pi =\sum_{i\,\in\, V} \pi_i\delta_i $ related to the stationary distribution $\pi$
with respect to which the Markov operator $T$
is {\it self-adjoint},
\begin{equation}
\label{s_a_analogue}
\widehat{T}=\,\frac 12
\left( \pi^{1/2}\,\, T\,\,
\pi^{-1/2}+\pi^{-1/2}\,\, T^\top\,\,
 \pi^{1/2}\right),
\end{equation}
where $T^\top$ is the adjoint operator. The orthonormal ordered set of real
 eigenvectors $\psi_i$, $i=1\ldots N$, of the symmetric operator $\widehat{T}$
 defines a basis in  $\mathcal{H}$. In the theory of
  random walks  defined on graphs \cite{Lovasz:1993,Aldous} and in spectral
  graph theory \cite{Chung:1997}, the properties of graphs are studied in
   connection with the  eigenvalues and eigenvectors of self-adjoint operators
   defined on them. In particular, the symmetric transition operator of the
random walk
 defined on
   undirected graphs is $\widehat{T_{ij}}=1/\sqrt{k_ik_j}$ if $i\sim j$. Its first eigenvector
    $\psi_1$ belonging to the largest eigenvalue $\mu_1=1$,
\begin{equation}
\label{psi_1}
\psi_1
\,\widehat{ T}\, =\,
\psi_1,
\quad \psi_{1,i}^2\,=\,\pi_i,
\end{equation}
describes the {\it local} property of nodes (connectivity), $\pi_i=k_i/2M,$
 where $2M=\sum_{i\in V} k_i$, while the remaining eigenvectors
 $\left\{\,\psi_s\,\right\}_{s=2}^N$ belonging to the eigenvalues
  $1>\mu_2\geq\ldots\mu_N\geq -1$ describe the {\it global} connectedness of the graph.

Markov's symmetric transition operator $\widehat{T}$  defines a {\it projection}
 of any density $\sigma\in \mathcal{H}$ on the eigenvector $\psi_1$ of the
  stationary distribution $\pi$,
\begin{equation}
\label{project}
\sigma\,\widehat{T}\,
=\,\psi_1 + \sigma^\bot\,\widehat{T},\quad \sigma^\bot\,=\,\sigma-\psi_1,
\end{equation}
in which $\sigma^{\bot}$ is the vector belonging to the orthogonal complement of
$\psi_1$.
Thus, it is clear that any two densities $\sigma,\rho\,\in\,\mathcal{H}$ differ
 with respect to random walks only by their dynamical components, $(\sigma-\rho)\,
 \widehat{T}^t\,=\,(\sigma^\bot -\rho^\bot)\,
\widehat{T}^t$ for all $t\,>\,0$.
Therefore, we can define the
distance  $\|\ldots\|_T$ between any two densities which they acquire  with
 respect to random walks by
\begin{equation}
\label{distance}
\left\|\,\sigma-\rho\,\right\|^2_T\, =
\, \sum_{t\,\geq\, 0}\, \left\langle\, \sigma-\rho\,\left|\,\widehat{T}^t\,
\right|\, \sigma-\rho\,\right\rangle.
\end{equation}
 or, using the spectral
representation of $\widehat{T}$,
\begin{equation}
\label{spectral_dist}
\left\|\,\sigma-\rho\,\right\|^2_T\, =
\, \sum_{t\,\geq 0}\, \sum_{s=2}^N\, \mu^t_s \,\left\langle\,
\sigma-\rho\,|\,\psi_s\right\rangle\!\left\langle\, \psi_s
\,|\, \sigma-\rho\,\right\rangle
 \, =\,  \sum_{s=2}^N\,\frac{\left\langle\, \sigma-\rho\,|\,
 \psi_s\right\rangle\!\left\langle\, \psi_s
\,|\, \sigma-\rho\,\right\rangle}{\,1\,-\,\mu_s\,},
\end{equation}
where we have used  Dirac's bra-ket notations especially
convenient in working with inner products and
rank-one
operators in Hilbert space.

If we introduce in $\mathcal{H}(V)$ a new inner product  by
\begin{equation}
\label{inner-product}
\left(\,\sigma,\rho\,\right)_{T}
\,= \, \sum_{t\,\geq\, 0}\, \sum_{s=2}^N
\,\frac{\,\left\langle\, \sigma\,|\,\psi_s\,\right\rangle\!
\left\langle\,\psi_s\,|\,\rho \right\rangle}{\,1\,-\,\mu_s\,}
\end{equation}
for all  $\sigma,\rho\,\in\, \mathcal{H}(V),$
then (\ref{spectral_dist}) is nothing else but
\begin{equation}
\label{spectr-dist2}
\left\|\,\sigma-\rho\,\right\|^2_T\, =
\left\|\,\sigma\,\right\|^2_T +
\left\|\,\rho\,\right\|^2_T  -
2 \left(\,\sigma,\rho\,\right)_T,
\end{equation}
 where
\begin{equation}
\label{sqaured_norm}
\left\|\, \sigma\,\right\|^2_T\,=\,
\,\sum_{s=2}^N \,\frac{\left\langle\, \sigma\,|\,\psi_s\,\right\rangle\!
\left\langle\,\psi_s\,|\,\sigma\, \right\rangle}{\,1\,-\,\mu_s\,}
\end{equation}
is the square
of the
 norm of  $\sigma\,\in\, \mathcal{H}(V)$ with respect to
random walks.
We finish the description of the $(N-1)$-dimensional Euclidean
space structure of $G$
induced by
  random walks by mentioning that
given two densities $\sigma,\rho\,\in\, \mathcal{H}(V),$ the
angle between them can be introduced in the standard way,
\begin{equation}
\label{angle}
\cos \,\angle \left(\rho,\sigma\right)=
\frac{\,\left(\,\sigma,\rho\,\right)_T\,}
{\left\|\,\sigma\,\right\|_T\,\left\|\,\rho\,\right\|_T}.
\end{equation}
Random walks embed connected undirected graphs into Euclidean
space. This embedding  can be used in order to compare
 nodes
and to retrace
 the optimal coarse-graining
representations.
Namely,  let us consider
the density $\delta_i$ which equals 1 at
the node $i\,\in\, V$ and zero for all other nodes.
With respect to the measure
$m_\pi$, it corresponds to the density
$\upsilon_i\,=\,\pi^{-1/2}_i\,\delta_i$.
Then, the square of the norm of  $\upsilon_i$ is given by
\begin{equation}
\label{norm_node}
\left\|\,\upsilon_i\,\right\|_T^2\, =\,{\frac 1{\pi_i}\,\sum_{s=2}^N\,
\frac{\,\psi^2_{s,i}\,}{\,1-\mu_s\,}},
\end{equation}
where $\psi_{s,i}$ is the $i^{\mathrm{th}}$-component of the
eigenvector $\psi_s$. In the theory of random walks \cite{Lovasz:1993},
the quantity (\ref{norm_node}) expresses the {\it access time} to a target node
quantifying the expected number
of  steps
required for a random walker
to reach the node
$i\in V$ starting from an
arbitrary
node  chosen randomly
among all other
nodes  with respect to
the stationary distribution $\pi$.

The Euclidean distance between any two nodes of the graph $G$
induced by random walks,
\begin{equation}
\label{commute}
K_{i,j}\,=\,\left\|\, \upsilon_i-\upsilon_j\,\right\|^2_T,
\end{equation}
is simply the {\it commute times} in theory of random walks and
is equal to the expected number of steps required for a random
walker starting at $i\,\in\, V$ to visit $j\,\in\, V$ and then to
return to $i$ again,  \cite{Lovasz:1993}.

It is important to mention that
the cosine of an angle calculated in accordance to
 (\ref{angle}) has the structure of
Pearson's coefficient of linear correlations
 that reveals it's natural
statistical interpretation.
Correlation properties of flows
of random walkers
passing by different paths
 have been remained beyond the scope of
previous  studies devoted to complex
networks and random walks on graphs.
The notion of angle between any two nodes in the
graph arises naturally as soon as we
become interested in
the strength and direction of
a linear relationship between
two random variables,
the flows of random walks moving through them.
If the cosine of an angle (\ref{angle}) is 1
(zero angles),
there is an increasing linear relationship
between the flows of random walks through both nodes.
Otherwise, if it is close to -1 ($\pi$ angle),
  there is
a decreasing linear relationship.
The  correlation is 0 ($\pi/2$ angle)
if the variables are linearly independent.
It is important to mention that
 as usual the correlation between nodes
does not necessary imply a direct causal
relationship (an immediate connection)
between them.

\subsection{Graph partitioning by random walks}
\label{subsec:Graph_Partitioning}
\noindent

Visual segmentation of networks based on 3D representations of
 their dual graphs is not always feasible.
The graph partitioning problem seeks to partition a weighted undirected
graph $G$ into $n$ weakly connected components $\Gamma_1,\ldots\Gamma_n$ such that
 $\bigcup_{i=1}^n\Gamma_n\subset G$ and either their properties share some common trait or
the graphs nodes belonging to them are close to each other according to some distance
 measure defined on nodes of the graph. A number of different graph
 partitioning strategies for undirected weighted  graphs
 have been studied in  connection with object recognition and learning in
  computer vision \cite{Vision}. In statistics, Principal Component
   Analysis (PCA) is used for the reducing size
 of a data set.
It is achieved by the optimal linear transformation retaining
the subspace that has largest variance (a lower-order principal component)
 and ignoring higher-order ones \cite{PCA,PCA2}.

Given an operator $S$ self-adjoint with respect
to  the  measure $m$
defined on a connected undirected graph $G$,
it is well known
  that the eigenvectors of the symmetric matrix ${\bf S}$
form an ordered orthonormal basis $\left\{\phi_k\right\}$
  with real eigenvalues
$\mu _1\geq\ldots\geq \mu_N$.
The ordered orthogonal basis
 represents the directions of the
 variances of variables
described by $S$. The number of
components which
may be detected in a network
with regard to
a certain dynamical process defined on it
depends upon the number of {\it essential
eigenvectors} of $S$.

Let us consider the normalized Laplace operator,  \cite{Chung:1997},
\begin{equation}
\label{norm_Lapalce}
\widehat{ L_{ij}}\,=\,\delta_{ij}-\widehat{T_{ij}},
\end{equation}
where $\widehat{T_{ij}}$ is the symmetric Markov transition operator $\widehat{T_{ij}}$.
Its  eigenvalues,
\begin{equation}
\label{tau_lambda}
0\,=\,\lambda_1\,<\,\ldots\,\leq\,\lambda_n\,\leq\,\tau^{\,-1}\,<\,\ldots\,\leq
\,\lambda_N\,\leq\,2,
\end{equation}
 can be interpreted as
the inverse characteristic time scales
of the diffusion
  process such that the smallest eigenvalues
correspond to the stationary distribution $\pi$
 together with the slowest diffusion
   modes involving the most
significant amounts of flowing commodity.

There is
 a simple time scale argument
which we use in order to determine the number of
applicable eigenvectors from the ordered orthogonal basis of eigenvectors,
 $\left[\,{\bf f}_1,\ldots {\bf f}_N\,\right]$.
It is obvious that while observing the network
 close to an equilibrium state
during short time, we detect flows resulting from
 a large number of transient  processes
evolving toward the stationary  distribution
and being characterized by the relaxation times
 $\propto\lambda^{-1}_k$.
While measuring the flows in
sufficiently long time $\tau$,
we may discover just $n$ different
eigenmodes (\ref{tau_lambda}).
In general, the longer is the time of measurements $\tau$,
the less is the
number of eigenvectors we
have to take into account in network component
 analysis of the network.
Should the time of measurements  be fixed, we can
determine the number $n$ of required eigenvectors.

In order to obtain the best quality segmentation,
it is convenient to center the $n$ primary eigenvectors.
The {\it centroid} vector
(representing the center of mass of
the set $\left[\,{\bf f}_1,\ldots {\bf f}_n\,\right]$)
is calculated as
the arithmetic mean,
\begin{equation}
\label{centroid_f}
{\bf m}\,=\,\frac 1n \,\sum_{k=1}^{n}\, {\bf f}_{k}.
\end{equation}
Let us denote the $n\,\times\, N$ matrix of $n$
centered eigenvectors  by
\[
{\bf F}\,=\,\left[\,{\bf f}_1\,-\,{\bf m},\ldots \,{\bf f}_N\,-\,{\bf m}\right].
\]
Then, the symmetric matrix of {\it covariances}
 between the entries of eigenvectors $\{{\bf f}_k\}$
is the
product of ${\bf F}$ and its adjoint ${\bf F}^\top$,
\begin{equation}
\label{Covariance}
\mathrm{\bf Cov}\,=\,\frac {\,\,{\bf F}\,{\bf F}^\top\,}{N\,-\,1}
\end{equation}
It is important to note that the
correspondent
 Gram matrix ${\bf F}^\top\,{\bf F}\,/(N-1) \,\equiv\, {\bf 1}$
due to
the orthogonality of the basis eigenvectors.
The main contributions in
the symmetric matrix $\mathrm{\bf Cov}$ are related
to the groups of nodes
\begin{equation}
\label{directions}
\mathrm{\bf Cov}{\ }{\bf u}_k\,=\,\sigma_k\,{\bf u}_k,
\end{equation}
 which can be identified by means of the
 eigenvectors
$\{\,{\bf u}_k\,\}$ associated to the first largest
eigenvalues among
 $\sigma_1\,\geq\, \sigma_2,$ $\dots,\,\geq \,\sigma_N$.
By ordering the eigenvectors  in decreasing order
 (largest first), we generate
an ordered orthogonal basis with the
  first eigenvector having the direction of largest variance of the
   components of  $n$ eigenvectors $\{{\bf f}_k\}$.
Let us note that
    due to the structure of ${\bf F}$ only the first $n-1$
    eigenvalues $\sigma_k$ are not trivial.
In accordance to the standard PCA notation,
the eigenvectors of the covariance
matrix ${\bf u}_k$  are called the
{\it principal directions} of the network
with respect to the diffusion process defined by
the operator $S$. A low dimensional representation of the network
is given by its principal directions
$\left[\,{\bf u}_1,\ldots,
{\bf u}_{n-1}\,\right],$ for $n\,<\,N$.

Diagonal
elements of the matrix $\mathrm{\bf Cov}$
quantify
the component variances of the eigenvectors
$\left[\,{\bf f}_1,\ldots {\bf f}_n\,\right]$
around their mean
 values (\ref{centroid_f}) and may be large
essentially for large networks. Therefore,
 it is practical for us
to use
the standardized {\it correlation} matrix,
\begin{equation}
\label{Corr}
{\mathrm Corr}_{ij}\,=\,\frac{\mathrm{Cov}_{ij}}{\sqrt{\mathrm{Cov}_{ii}}
\sqrt{\mathrm{Cov}_{jj}}},
\end{equation}
instead of the covariance matrix
${\mathrm{\bf Cov}}$.
We emphasize that
the diagonal elements of (\ref{Corr})
equal 1, while the off-diagonal
elements are the Pearson's
 coefficients of
linear correlations, \cite{Pearson}.

Let ${\bf U}$ be the orthonormal matrix which contains
the eigenvectors $\{{\bf u}_k\}$, $k\,=\,1,\ldots,\, n-1,$
of the covariance (or correlation) matrix
as the row vectors. These vectors form the orthogonal
 basis of the $(n-1)$-dimensional vector space, in
which every variance $({\bf f}_k-{\bf m})$ is
represented by a  point ${\bf g}_k\,\in\,\mathbb{R}^{\,(n-1)}$,
\begin{equation}
\label{cov_transform}
{\bf g}_k\,=\,{\bf U}\,({\bf f}_k-{\bf m}\,).
\end{equation}
Then each original eigenvector
   ${\bf f}_k$ can be obtained from
${\bf g}_k\,\in\,\mathbb{R}^{\,(n-1)}$ by the inverse transformation,
\begin{equation}
\label{inverse_transff}
{\bf f}_k\,=\,{\bf U}^\top\,{\bf g}_k\,+\,{\bf m}.
\end{equation}
Using transformations (\ref{cov_transform}) and (\ref{inverse_transff})
we obtain the $(n-1)$-dimensional representation
$\left\{\,\varphi_k \,\right\}_{\,k=1}^{\,(n-1)}$
of the $N$-dimensional basis vectors
$\left\{\, {\bf f}_s\,\right\}_{\,s=1}^{\,N}$
in the form
\begin{equation}
\label{compress}
\varphi_{\,k}\,=\,{\bf U}^\top{\bf U}\,{\bf f}_k \,+\,
\left(\,{\bf 1}-{\bf U}^\top{\bf U}\,\right)\,{\bf m},
\end{equation}
that minimizes the mean-square error between
 ${\bf f}_k\,\in\,\mathbb{R}^{\,N}$
and $\varphi_k\,\in\,\mathbb{R}^{\,(n-1)}$ for given $n$.

Variances of eigenvectors $\left\{\,{\bf f}_k\,\right\}$
are positively correlated within a principal component
 of the transport network. Thus, the
transition matrix ${\bf U}^{\top}{\bf U}$
can be interpreted as
the connectivity patterns acquired
by the network with respect to the diffusion process.
Two nodes, $i$ and $j$,
belong to one and the same principal
component  of the network
if $\left({\bf U}{\bf U}^\top\right)_{ij}\,>\,0$.
By applying the Heaviside function, which is
zero for negative argument and one for positive argument,
to the elements of the transition matrix
${\bf U}{\bf U}^\top$, we derive the
coarse-grained
connectivity matrix of network components.

\section{Thermodynamics of Transport Networks}
\label{sec:Thermodynamics}
 \noindent

In the present section, we study the spectrum (\ref{tau_lambda})
of the normalized Laplace operator (\ref{norm_Lapalce}) defined
on a connected, undirected graph $G$ by means of spectral methods
strongly inspired by statistical mechanics.  The obvious advantage
 of statistical mechanics  is that statistical moments of random
 walks would acquire a "physical interpretation" in the framework
 of this thermodynamic formalism.
The full description of transport
networks in short time and small scales requires a high
dimensional space, e.g. the knowledge of locations and velocities
of all agents participating in transport processes. Being
defined on a graph, the time evolution of such a system can be
described by just a few dynamically relevant variables called
reaction coordinates \cite{Lafon}. Identification of
slow variables and dynamically meaningful reaction coordinates that
capture the long time evolution of transport systems is among the most important
problems of transport network analysis. The spectrum of Laplace
   operator could have gaps indicating the time scales separation,
    that is, there are only a few "slow" time scales at which the
    transport network is meta-stable, with many "fast" modes
    describing the transient processes toward the slow modes.
The methods of spectral graph theory allow detecting and separating
"slow" and "fast" time scales giving rise to the component
analysis of networks in which the primary eigenvalues play the
essential role.

\subsection{Spectral function of Laplace operator}
\label{subsec:Spectral_Function}
\noindent

Since the spectra of
 self-adjoint Laplace operators defined on the undirected graphs are
 non-negative they can be investigated by means of the
characteristic functions which discriminate contributions from the
largest eigenvalues of Laplace operator in favor of those from the
minimal ones. These characteristic functions are usually
associated with kernels.

Given a self-adjoint operator $L$ defined on a finite dimensional Hilbert space
$\mathcal{H}$ the latter has an
orthonormal basis $\{\phi_k\}_{k=1}^N$  and every element of the Hilbert space
 can be written in a unique way as a sum of multiples of these basis elements.
The use of Borel's functional calculus    for a positive semi-definite
 operator function $f(L)$ allows us to define a Hilbert space $\mathcal{H}$ on
 $\mathbb{R}^N$ via the dot product,
\begin{equation}
\label{dot_product}
\left\langle
\phi,\phi
\right\rangle_{\mathcal{H}}
=\left\langle
\phi,f(L)\phi
\right\rangle,
\end{equation}
with the {\it kernel} $f^{-1}(L)$. The kernels associated with
the Laplace operator defined on an undirected graph $G$ play the role of the
{\it Green functions}
 describing long-range interactions between different
diffusion modes induced
 by the graph structure (the main graph components).
The Green function recently introduced in \cite{Kondor_Laff} is specified by
the parameter $\beta>0$,
\begin{equation}
\label{heat_kernel}
\left({\bf K}_\beta\right)_{ij}=\exp(-\beta L_{i,j}),
\end{equation}
and solve the diffusion equation defined on a connected, undirected  graph $G$:
\begin{equation}
\label{solution_heat}
\mathcal{G}=\int_0^\infty {\bf K}_\beta {\ }d\beta=
\int_0^\infty\exp(-\beta {\bf L}){\ }d\beta=
{\bf L}^{-1},
\end{equation}
 with initial conditions being a
source at node $i\in V$ at $\beta=0$ \cite{Zhu}.
The function $\left({\bf K}_\beta\right)_{ij}$
 describes the expected number of particles that would accumulate at
 vertex $j$ after a given amount of time if they have been injected
 at vertex $i$ and diffused through the graph along the edges.
The exponential diffusion kernel (\ref{heat_kernel}) discriminate
large eigenvalues that is preferable while performing estimation
on the graph since it
 biases the estimate towards functions which vary little over the graph
  components \cite{Saitoh}-\cite{Wahba}.

The distribution of eigenvalues $\lambda_k\in[0,2[$ of the self-adjoint Laplace operator
(\ref{norm_Lapalce}) defined on the undirected graph $G$
are characterized by
the {\it spectral density function},
\begin{equation}
\label{rho}
\rho(\lambda)=\sum_{k=1}^N\delta(\lambda-\lambda_k).
\end{equation}
However, it is more convenient to study the properties of non-negative spectra by means of
 the {\it characteristic functions}.

The first characteristic function is given by the Laplace transform of the spectral
 density (\ref{rho}),
\begin{equation}
\label{Laplace_transform}
\begin{array}{lcl}
Z(\beta)&=& \int_0^{\infty}e^{-\beta\lambda}\rho(\lambda){\ }d\lambda\\
 & =& \sum_{k=1}^N e^{-\beta \lambda_k},
\end{array}
\end{equation}
known as the {\it canonical partition function}.
 encoding the statistical properties of the system being
in the thermodynamic equilibrium.
The {\it inverse temperature} parameter
$\beta>0$ which can be considered either as an effective time
scale in the problem (the number of spaces in the dual graph a walker passes through in
one time step) or as the laziness parameter.

\subsection{Thermodynamics of urban networks}
\label{subsec:Thermodynamic_potentials}
\noindent

Statistical mechanics provides a framework for relating the microscopic
properties of a system to its macroscopic or bulk properties.
The statistical properties of diffusions
defined on undirected graphs can be described by means of canonical partition function
(\ref{Laplace_transform}),
in which $\beta^{-1}$ plays the role of the inverse temperature.
In statistical mechanics, the temperature  arises from the fact that the system is
in the thermal equilibrium with its environment that is achieved when two systems
in thermal contact with each other cease to exchange energy by heat \cite{StatMech}.
Equilibrium implies a state of balance and can be naturally interpreted as the
 stationary distribution $\pi$ in the framework of random walks defined on the
 undirected graphs.

Bringing up the concept of the canonical ensemble it is possible to derive
the probability $P_i$ that random walkers
will be in  a certain "microstate" characterized by the spectral
value $\lambda_i$:
\begin{equation}
\label{microstate}
P_i=Z^{-1}(\beta)\exp(-\beta\lambda_i).
\end{equation}
 Therefore, the partition function (\ref{Laplace_transform}) can
 be used to find the expected value of any microscopic
  properties of flows of the random walkers, which can
   then be related to some  macroscopic variables
    related to the entire system. It is worth to mention
     that since random walkers neither have masses nor
     kinetic energy, the nontrivial macroscopic variables
      that can be derived from the partition function
      (\ref{Laplace_transform}) characterize the
the certain structural properties of the graph with
 respect to the self-adjoint operator defined on it.

We have discussed the thermodynamic potentials pertinent to space syntax in
five different compact urban patterns in \cite{Volchenkov:2007}.
Two of them are
situated on islands: Manhattan (with an almost regular grid-like
city plan) and the network of Venice canals
(Venice stretches across 122 small islands, and
 the canals serve the function of roads).
We have also considered two organic cities
founded shortly after the Crusades and developed within the medieval
fortresses: Rothenburg ob der Tauber (the medieval Bavarian city
preserving its original structure from the 13$^\mathrm{th}$ century)
and the downtown of
Bielefeld  (Altstadt Bielefeld),
 an economic and cultural center of Eastern Westphalia.
To supplement the study of urban canal networks, we have
investigated that one in the city of Amsterdam. We do not
reproduce the results reported in \cite{Volchenkov:2007} in the
present paper refereing interested readers to
\cite{Volchenkov:2007} for details. Below, we just give an outline
of the approach and study the thermodynamic parameters which have
not been investigated in our previous work.

\begin{figure}[ht]
 \noindent
\begin{center}
\epsfig{file=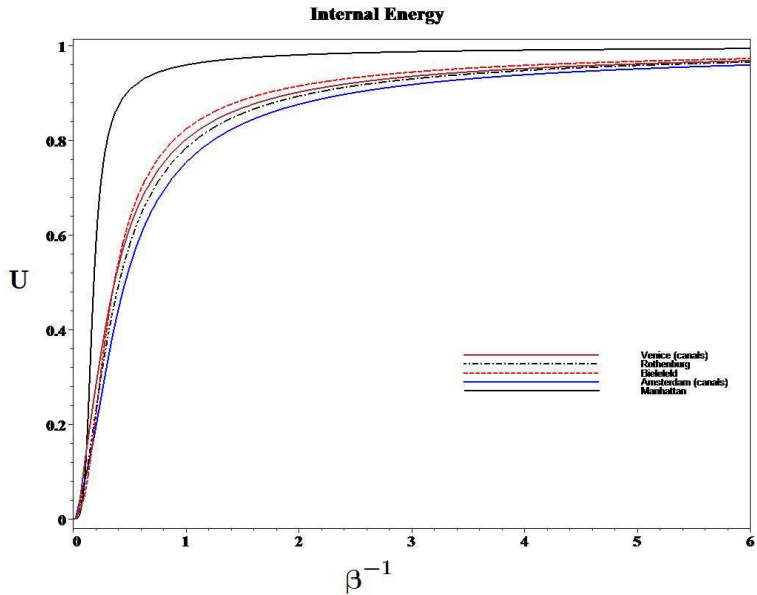, angle= 0,width =10cm, height =8cm}
  \end{center}
\caption{\small The expected values of the "microscopic energy"
(the averaged eigenvalues) are calculated for the spectra of compact
 urban patterns.
The slopes of internal energy curve for the highly regular street
grid of Manhattan is steeper than those for
 the street patterns of the organic cities (Bielefeld, Rothenburg,
  canals of Amsterdam and Venice).}
 \label{Fig2_12}
\end{figure}

\begin{figure}[ht]
 \noindent
\begin{center}
\epsfig{file=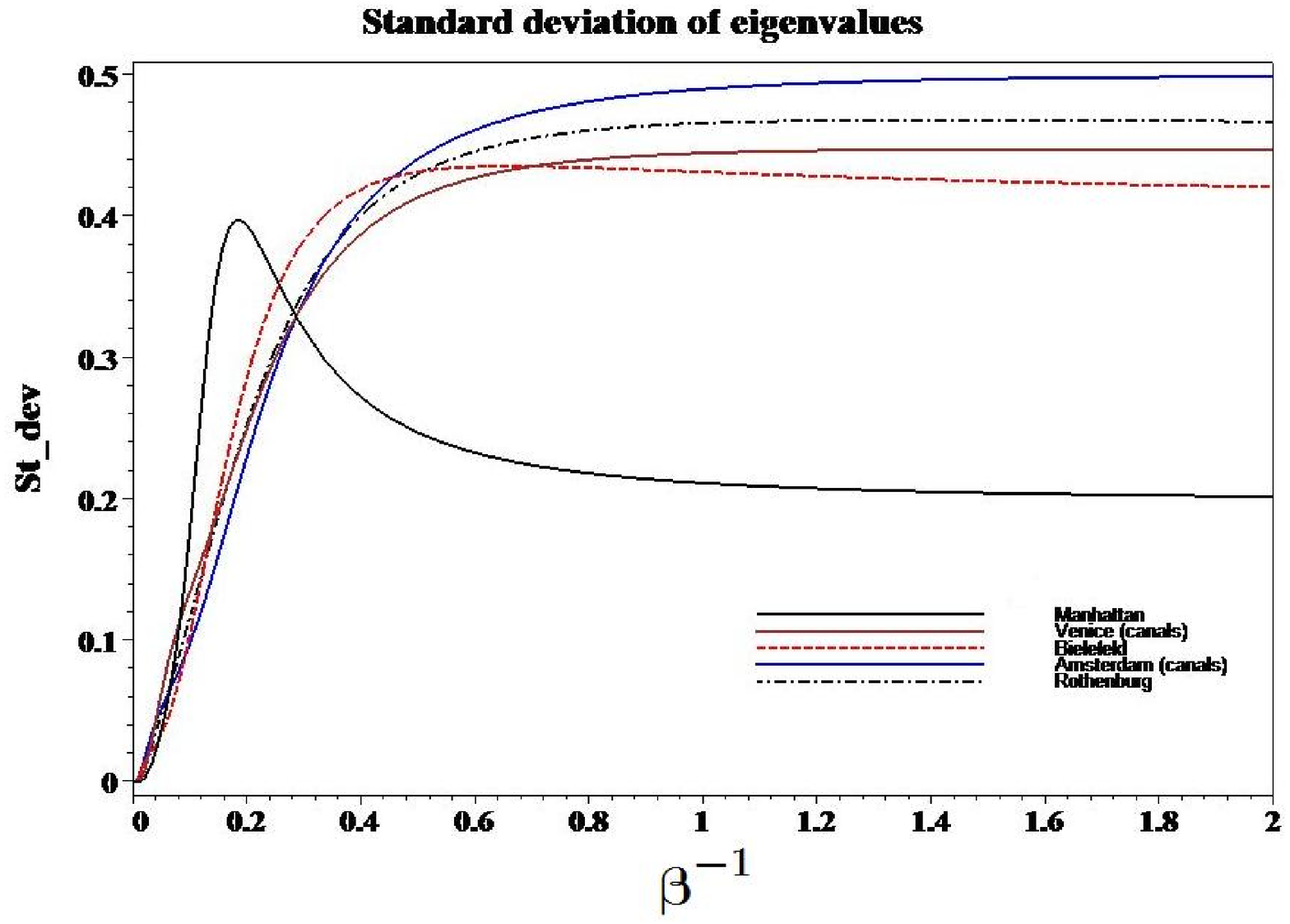, angle= 0,width =10cm, height =8cm}
  \end{center}
\caption{\small The deviations of eigenvalues around the "microscopic energy"
(the averaged eigenvalues) are calculated for the spectra of compact
 urban patterns.}
 \label{Fig2_12a}
\end{figure}

\begin{figure}[ht]
 \noindent
\begin{center}
\epsfig{file=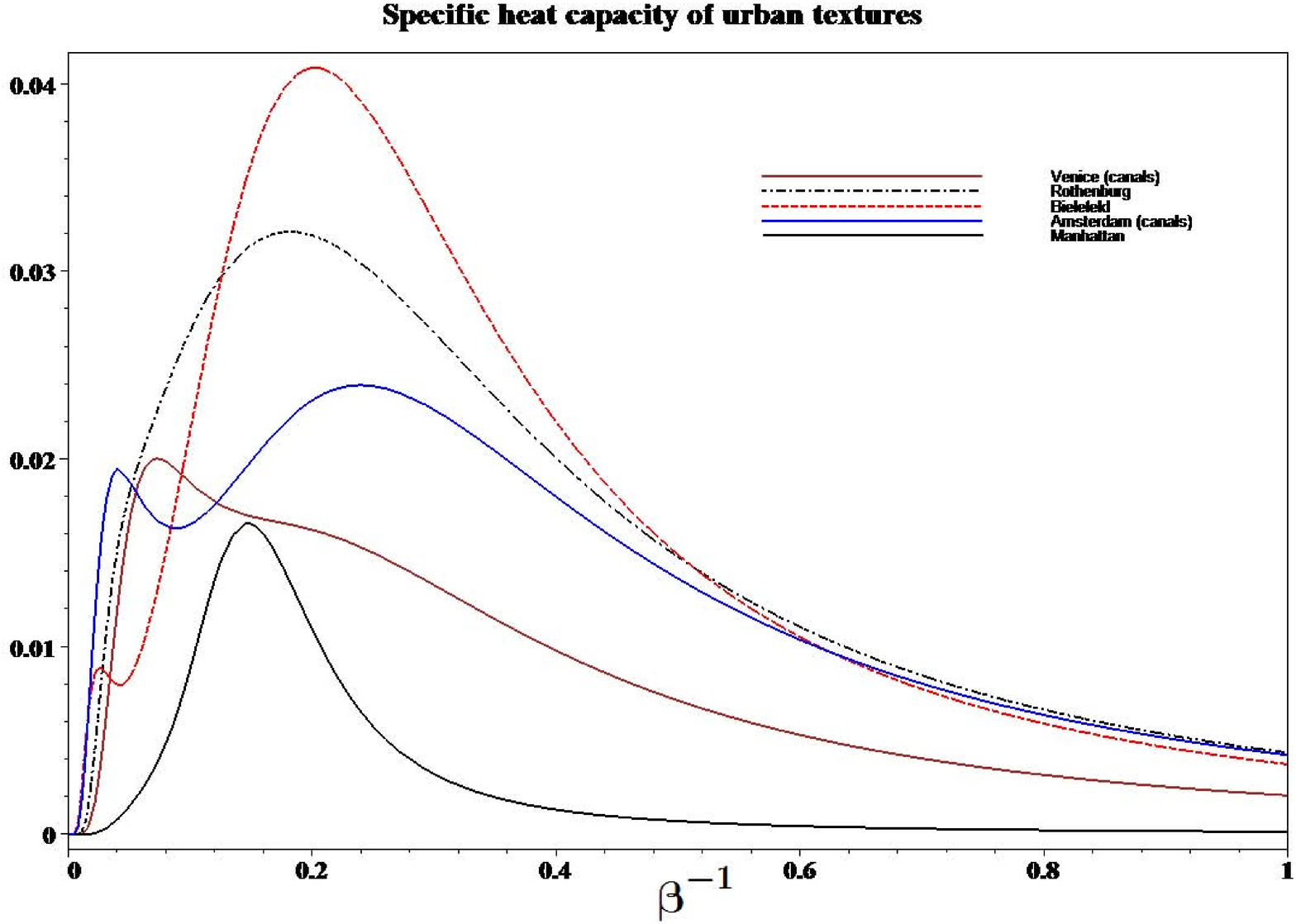, angle= 0,width =10cm, height =8cm}
  \end{center}
\caption{\small The specific heat capacity, as a function of
temperature in the model of random walks defined on the different urban textures. }
 \label{Fig2_18}
\end{figure}

The expected value of the "microscopic energy" (the
 averaged eigenvalue) held at constant temperature $\beta^{-1}$,
\begin{equation}
\label{internal_energy}
\langle \lambda\rangle
\equiv U=
\frac 1{Z(\beta)}\sum_{i=1}^{N}\lambda_i e^{-\beta\lambda_i}=
-\frac{\partial_\beta Z(\beta)}{Z(\beta)}=-\partial_\beta\ln Z(\beta),
\end{equation}
can be interpreted as the microscopic definition of the
 thermodynamic variable
corresponding to the
 internal energy in statistical
  mechanics. Due to the complicated topology of streets and canals, the flows
of random walkers exhibit  spectral properties similar to that
of a thermodynamic system characterized by a nontrivial internal
energy (see Fig.~\ref{Fig2_12}).
In Fig.~\ref{Fig2_12}, one can clearly see the difference between the
street patterns of the organic cities (Bielefeld, Rothenburg, canals
of Amsterdam and Venice) and the street grid of Manhattan.
Taking the derivative of $\langle \lambda\rangle$ with respect to the parameter
$\beta$ in (\ref{internal_energy}), we have
\begin{equation}
\label{variance}
\frac{d\langle \lambda\rangle}{d\beta}=
\frac 1{Z^2(\beta)}\left(\frac{dZ(\beta)}{d\beta}\right)^2- \frac 1{Z(\beta)}\left(
\frac{d^2Z(\beta)}{d\beta^2}\right) =
\langle\lambda\rangle^2-\langle\lambda^2\rangle
= -D^2(\lambda),
\end{equation}
where $D^2(\lambda)$ is the {\it variance}, the measure of its
statistical dispersion, indicating how the eigenvalues of the
normalized Laplace operator are spread around the expected value
$U$ indicating  the variability of the eigenvalues. The standard
form of the usual deviations from the mean is the standard
deviation,
\begin{equation}
\label{st_dev}
\sigma=\sqrt{D^2(\lambda)}.
\end{equation}
A large standard deviation indicates that the data points
are far from the mean and a small standard deviation
indicates that they are clustered closely around the mean (see Fig.~\ref{Fig2_12a}).
In thermodynamics, heat capacity is the rate of change of
temperature as heat is added to a body at the given conditions
and state of the body (foremost its temperature). Similarly to
 pressure, it is an extensive quantity being sensitive to the
 network size.
Dividing heat capacity by the network size yields a specific
 heat capacity, which is an intensive quantity that is no longer
 dependent on $N$, and is now dependent on the graph structure,
\begin{equation}
\label{specificheatC}
{\mathcal C}=\frac 1N \left(\frac{\partial U}{\partial T}\right)
\end{equation}
We have presented the specific heat capacity, as a function of
temperature in the model of random walks defined on the different
 compact cities in Fig.~\ref{Fig2_18}.

Note that the specific heat capacity is zero at zero temperature
($\beta\to \infty$), and rises to a maximal value as temperature
grows up. It is interesting to note that the specific heat
capacity in the random walks models defined on networks is
dramatically different from that of most physical systems (like
the multi-atomic gases or a crystalline solid phase); it is not a
monotonous increasing function approaching the  Dulong-Petit limit
at higher temperature (the theoretical maximum of heat capacity).
The decay of specific heat at high temperatures ($\beta\to 0$)
results from the "freezing out" of some degrees of freedom as
$\beta\to 0$ (when random walkers rest at nodes most of the time).

The maximal values  of the specific heat capacity is an individual
 characteristics of a city structure. They are higher for the
 organic cities and lower for those planned in a regular grid.

\section{Directed and Interacting Transport Networks}
\label{sec:Directed}
 \noindent

Many transport networks correspond to {\it directed} graphs.
Power grids transporting electrical currents and driving
directions assigned to city streets in order to optimize traffic
in a city are the examples \cite{AlbertBarabasi2002}.
Traffic
within large cities is formally organized with marked driving directions
creating one-way streets.
On those streets all traffic must flow in only one direction, but
pedestrians on the sidewalks are generally not limited to one-way movement.
It is well known that the use of one-way streets would
greatly improve traffic flow since
the speed of traffic is increased and intersections are
simplified.

Moreover, even if the network of city itineraries constitutes an
undirected graph, there is a general principle that establishes
who has the right to go first while crossing the road
intersections or other conflicting parts of the road and who has
to wait until the other driver does so. While assigning weights
estimating traffic loading of itineraries on such a city graph, we
could shortly discover that each road side should be characterized
by its own weight depending upon the number of right-hand
junctions the road has with other itineraries that is a directed weighted graph.

The spectral approach for directed graphs has not been as well
developed as for undirected graph. Indeed, it is rather
difficult if ever possible to define a {\it unique} self-adjoint
operator on directed graphs. In general, any node $i$ in a
directed graph $\vec{G}$ can have different number of
in-neighbors and out-neighbors,
\begin{equation}
\label{deg_in_out}
k_\mathrm{in}(i)\,\ne\, k_\mathrm{out} (i).
\end{equation}
In particular, a node $i$ is a {\it source} if
$k_\mathrm{in}(i)=0, $ $k_\mathrm{out}(i)\ne 0$,
and is a {\it sink} if $k_\mathrm{out}(i)=0$,
$k_\mathrm{in}(i)\ne 0$. If the graph has neither  sources nor sinks,
it is called  strongly connected.

Despite asymmetric interactions represented by directed graphs are
ubiquitous in many technological networks, they have not been yet very
well studied in complex network theory.
In general, the local structure of directed graphs is fundamentally
different from that of undirected graphs.
In particular, the diameters of directed networks can essentially
 exceeds that the one corresponding to the same networks regarded as undirected.
A recent investigation into the local structure of directed
networks \cite{Bianconi} shows that directed networks often have
very few short loops as compared to random models usual in
contemporary theory of complex systems.
In undirected networks, the high density of short loops (high
clustering coefficient) together with small graph diameter gives
rise to the small-world effect \cite{WattsStrogatz}. In directed
networks, the correlation between number of incoming and outgoing edges
 modulates the expected number of short loops. In particular,
it has been demonstrated in \cite{Bianconi} that if the values
$k_{\mathrm{in}}(i)$ and $k_{\mathrm{out}}(i)$ are not
correlated, then the number of short loops is strongly reduced as
compared to the case when both degrees are positively correlated.

\subsection{Random walks on directed graphs}
\label{subsec:RW_on_directed_graphs}
\noindent

Finite random walks
are defined on a strongly connected directed graph $\vec{G}(V,\vec{E})$
 as finite vertex sequences
$\mathfrak{w}=\{v_0,\ldots,v_n\}$ (time forward) and
$\mathfrak{w}'=\{v_{-n},\ldots,v_0\}$ (time backward)
such that each pair
$(v_{i-1},v_i)$ of vertices adjacent either in $\mathfrak{w}$
 or in  $\mathfrak{w}'$ constitutes a directed edge
$v_{i-1}\to v_i$ in $\vec{G}$.

A {\it time forward} random walk is defined by the transition probability
matrix \cite{Chung2005} $P_{ij}$
for each pair of nodes $i,j\in \vec{G}$ by
\begin{equation}
\label{trans_directed}
P_{ij}=\left\{
\begin{array}{ll}
 1/k_{\mathrm{out}(i)}, & i\to j, \\
0, & \mathrm{otherwise},
\end{array}
\right.
\end{equation}
which satisfies the
 probability conservation property:
 \begin{equation}
\label{probab_conserv1}
\sum_{j,{\ } i\to j}P_{ij} =1.
\end{equation}
The definition (\ref{trans_directed}) can be naturally extended
for weighted graphs \cite{Chung2005} with $w_{ij}>0$,
\begin{equation}
\label{trans_weighted}
P_{ij}=\frac{w_{ij}}{\sum_{s\in V}w_{is}}.
\end{equation}
Matrices (\ref{trans_directed}) and (\ref{trans_weighted}) are
real, but not symmetric and therefore have complex conjugated
pairs of eigenvalues. For each pair of nodes $i,j\in\vec{G}$, the
{\it forward} transition probability is given by
$p^{(t)}_{ij}=({\bf P}^t)_{ij}$ that is equal zero, if  $\vec{G}$
does not
contain  a directed path from $i$ to $j$.

{\it Backwards time} random walks are defined on the strongly
connected directed graph $\vec{G}$ by the stochastic transition
matrix
\begin{equation}
\label{chip_firing}
{P}^\star_{ij}=
\left\{
\begin{array}{ll}
1/k_\mathrm{in}(i), & j\to i, \\
0, & \mathrm{otherwise},
\end{array}
\right.
\end{equation}
 satisfying
{\it another} probability conservation property
\begin{equation}
\label{probab_conserv2}
\sum_{i,{\ } i\to j}P^\star_{ij} =1.
\end{equation}
It describes random walks
unfolding backwards in time: should a random walker arrives at $t=0$
at a node $v_0$,
then
\begin{equation}
\label{backward_trans}
p^{(-t)}_{ij}=(({\bf P^\star})^{t})_{ij}
\end{equation}
defines
the probability that $t$ steps {\it before} it had originated from a node $j$.
The matrix element (\ref{backward_trans}) is zero, provided there is
no directed path from $j$ to $i $ in $\vec{G}$.

If $\vec{G}$ is strongly connected and {\it aperiodic}, the random walk converges
\cite{LovWinkl,Chung2005,chip-firing,chip-firing2} to
the {\it only} stationary distribution $\pi$ given by the Perron vector,
$\pi {\bf P}=\pi.$
If the graph $\vec{G}$ is periodic, then the transition
probability matrix ${\bf P}$ can have more than one eigenvalue
with absolute value 1, \cite{Chung:1997}. The components of Perron's vector $\pi$ can be
normalized in such a way that $\sum_{i}\pi_i=1$. The Perron vector  for
random walks defined on a strongly
connected directed graph can have coordinates with
exponentially small values, \cite{Chung2005}.

Given an aperiodic strongly connected graph $\vec{G}$, a
 self-adjoint Laplace operator
$L=L^{\top}$ can be defined \cite{Chung2005} as
\begin{equation}
\label{Laplace_directed}
L_{ij}\,=\,\delta_{ij}-\frac 12 \,\left(\pi^{1/2}{\bf P}\pi^{-1/2}+
\pi^{-1/2}{\bf P}^\top\pi^{1/2} \right)_{ij},
\end{equation}
where $\pi$ is a diagonal matrix with entries $\pi_i$. The matrix
(\ref{Laplace_directed}) is symmetric and has real non-negative
eigenvalues $0=\lambda_1<\lambda_2\leq\ldots\leq\lambda_N$ and
real eigenvectors.

It can be easily proven \cite{Butler2} that the
Laplace operator (\ref{Laplace_directed}) defined
on the aperiodic strongly connected graph $\vec{G}$
is equivalent to the Laplace operator defined on a
 {\it symmetric undirected weighted} graph $G_w$ on
the same vertex set with weights defined by
\begin{equation}
\label{weights_directed}
w_{ij}\, = \pi_iP_{ij}+\pi_jP_{ji}.
\end{equation}

\subsection{Bi-orthogonal decomposition of random walks defined
 on strongly connected directed graphs}
\label{subsec:Bi_orthogonal}
\noindent

In order to define the self-adjoint Laplace operator
(\ref{Laplace_directed}) on aperiodic strongly connected directed
graphs, we have to know the stationary distributions of random
walkers $\pi$. Even if $\pi$ exists for a given directed graph
$\vec{G}$, it can be evaluated usually only numerically in
polynomial time \cite{LovWinkl}.
 Stationary distributions on aperiodic
general directed graphs are
not so easy to describe since they are typically
{\it non-local} in  sense that
 each coordinate $\pi_i$
would depend upon the entire subgraph (the number of
spanning arborescences of
 $\vec{G}$ rooted at $i$ \cite{LovWinkl}), but
not on the local connectivity property of a node itself
like it was in undirected graphs. Furthermore, if the greatest common
divisor of its cycle lengths in
 $\vec{G}$ exceeds 1, then the transition probability matrices
  (\ref{trans_directed}) and (\ref{chip_firing})
can have several eigenvectors belonging to the largest eigenvalue
1, so that the definition
(\ref{Laplace_directed}) of Laplace operator seems to be questionable.

Given a strongly connected
directed graph $\vec{G}$
specified by the adjacency matrix
${\bf A}_{\vec{G}}$ $\ne {\bf A}^\top_{\vec{G}}$,
we consider two random walks operators.
A first transition operator
represented by the matrix
\begin{equation}
\label{Probab_out_directed}
{\bf P} = {\bf D}_\mathrm{out}^{-1}{\bf A}_{\vec{G}},
\end{equation}
 in which ${\bf D}_\mathrm{out}$ is a diagonal matrix
with entries $k_\mathrm{out}(i)$,
describes  the {\it time forward}
random walks of the nearest neighbor type defined
on $\vec{G}$. Given a time forward vertex sequence $\mathfrak{w}$
rooted at $i\in \vec{G}$, the matrix element
$P_{ij}$ gives the probability that $j\in\vec{G}$ is the vertex
next to $i$ in $\mathfrak{w}$.
A second transition operator, ${\bf P}^\star$, is dynamically conjugated
to (\ref{Probab_out_directed}),
\begin{equation}
\label{dynam_conjugated}
\begin{array}{lcl}
{\bf P}^\star &= &{\bf D}_\mathrm{in}^{-1}{\bf A}^\top_{\vec{G}}\\
              & = &   {\bf D}_\mathrm{in}^{-1}{\bf P}^\top{\bf D}_\mathrm{out},
\end{array}
\end{equation}
where ${\bf D}_\mathrm{in}$
is a diagonal matrix
with entries $k_\mathrm{in}(i)$.
The transition operator (\ref{dynam_conjugated}) describes random walks
over time backward vertex sequences $\mathfrak{w}'$.

It is worth to mention that   on undirected graphs
${\bf P}^\star \equiv{\bf P}$, since
$k_{\mathrm{in}}(i)=k_{\mathrm{out}}(i)$ for $\forall i\in\vec{G}$ and
${\bf A}_G={\bf A}^\top_G.$ While on directed graphs,
${\bf P}^\star$ is related to ${\bf P}$ by
the transformation
\begin{equation}
\label{conjugation}
{\bf P}={\bf D}_\mathrm{out}^{-1}\left({\bf P}^\star\right)^\top{\bf D}_\mathrm{in},
\end{equation}
so that these operators are not adjoint,
in general ${\bf P}^\top\ne {\bf P}^\star.$

We can define two different measures
\begin{equation}
\label{measures2}
\mu_+=\sum_j \deg_\mathrm{out}(j)\delta(j),\quad \mu_{-}=\sum_{j}\deg_\mathrm{in}(j)\delta(j)
\end{equation}
associated with the {\it out}- and {\it in-}degrees of nodes
of the directed graph.
In accordance to (\ref{measures2}), we define two
Hilbert spaces $\mathcal{H}_+$ and $\mathcal{H}_-$
corresponding to
the spaces of square summable functions,
$\ell^2\left(\mu_+\right)$ and  $\ell^2\left(\mu_-\right)$,
by setting the norms as
$$
\|x\|_{\mathcal{H}_\pm}=\sqrt{\left\langle x,x\right\rangle}_{\mathcal{H}_\pm},
$$
where $\left\langle \cdot,\cdot\right\rangle_{\mathcal{H}_\pm}$
 denotes the inner products with respect to measures (\ref{measures2}).
Then a function $f(j)$ defined on the set of graph vertices
is $f_{\mathcal{H}_-}(j)\in \mathcal{H}_-$ if transformed by
\begin{equation}
\label{transf_pm}
f_{\mathcal{H}_-}(j)\to J_- f(j) \equiv \mu_-{}_j^{-1/2} f(j)
\end{equation}
and  $f_{\mathcal{H}_+}(j)\in \mathcal{H}_+$ while being
transformed accordingly to
\begin{equation}
\label{transf_pm2}
f_{\mathcal{H}_+}(j)\to J_+ f(j) \equiv \mu_+{}_j^{-1/2} f(j).
\end{equation}
The obvious advantage of the measures (\ref{measures2}) against
the
 natural counting measure $\mu_0=\sum_{i\in V}\delta_i$
 is that the matrices of
the transition operators $P$ and $P^\star$ are
transformed under the change of measures as
\begin{equation}
\label{transf_op}
P_\mu=J_+^{-1}PJ_-, \quad P^\star_\mu=J_-^{-1}P^\star J_+,
\end{equation}
and become adjoint,
\begin{equation}
\label{adjoint}
\begin{array}{l}
\left(P_\mu\right)_{ij}=\frac{A_{\vec{G}}{}_{ij}}
{\sqrt{k_\mathrm{out}(i)}\sqrt{k_\mathrm{in}(j)}},
\\
\left(P^\star_\mu\right)_{ij}\equiv \left(P^\top_\mu\right)_{ij}=
\frac{A^\top_{\vec{G}}{}_{ij}}
{\sqrt{k_\mathrm{in}(i)}\sqrt{k_\mathrm{out}(j)}}.
\end{array}
\end{equation}
It is also important to note that
$$
P_\mu:\mathcal{H}_-\to \mathcal{H}_+
\quad \mathrm{and}
\quad P^\star_\mu:\mathcal{H}_+\to \mathcal{H}_-.
$$
We obtain the singular value dyadic expansion ({\it biorthogonal decomposition}
introduced in
 \cite{Lima1991,Aubry1991}) for the transition operator:
\begin{equation}
\label{biorthogonal}
{\bf P}_\mu=\sum_{s=1}^N\Lambda_s\varphi_{si}\psi_{si}\equiv
\sum_{s=1}^N\Lambda_s\left|\varphi_s\rangle\langle\psi_s\right|,
\end{equation}
where $0\leq \Lambda_1\leq \ldots \Lambda_N$ and
the functions $\varphi_k \in \mathcal{H}_+$ and $\psi_k\in \mathcal{H}_-$ are
related by  the
Karhunen-Lo\`{e}ve {\it dispersion} \cite{Karhunen,Loeve},
\begin{equation}
\label{Karhuen}
{\bf P}_\mu\varphi_s=\Lambda_s\psi_s
\end{equation}
satisfying the orthogonality condition:
\begin{equation}
\label{otrtho}
\langle \varphi_s,\varphi_{s'}\rangle_{\mathcal{H}_+}
=\langle \psi_{s'},\psi_s\rangle_{\mathcal{H}_-}=\delta_{s'\,s}.
\end{equation}
Since the operators $P_\mu$ and $P_\mu^\top$ act between different Hilbert
 spaces, it is
  insufficient to solve just one equation in
   order to determine their eigenvectors $\varphi_s$ and $\psi_s$, \cite{Lima_Aubry93}.
Instead, two equations have to be solved,
\begin{equation}
\label{two_eqs}
\left\{
\begin{array}{lcl}
{\bf P}_\mu\varphi&=&\Lambda\psi,\\
{\bf P}^\top_\mu\psi&=&\Lambda\varphi,
\end{array}
\right.
\end{equation}
or, equivalently,
\begin{equation}
\label{matrix_two_eqs}
\left(
\begin{array}{cc}
0 & {\bf P}_\mu \\
{\bf P}_\mu^\top& 0
\end{array}
\right)
\left(
\begin{array}{c}
\varphi \\
\psi
\end{array}
\right)
=\Lambda
\left(
\begin{array}{c}
\varphi \\
\psi
\end{array}
\right).
\end{equation}
The latter equation allows for a graph-theoretical interpretation.
The block anti-diagonal operator matrix in the left hand side of
(\ref{matrix_two_eqs}) describes random walks defined on a
 bipartite graph.
Bipartite graphs contain
two disjoint sets of vertices such that no edge has
both end-points in the same set.
However, in (\ref{matrix_two_eqs}), both sets are
formed by one and the same nodes of the
original graph $\vec{G}$
on which
two different random walk processes specified by the operators
$P_\mu$ and $P_\mu^\top$ are
defined.

It is obvious that any solution of the
equation (\ref{matrix_two_eqs}) is also a
 solution of the system
\begin{equation}
U\varphi=\Lambda^2\varphi, \quad
V\psi =\Lambda^2\psi,
\label{system01}
\end{equation}
in which $U\equiv {\bf P}_\mu^\top{\bf P}_\mu$ and
$V\equiv {\bf P}_\mu{\bf P}_\mu^\top$,
although the converse is not necessarily true.
The self-adjoint  nonnegative operators $U:\mathcal{H}_-\to\mathcal{H}_-$ and
$V:\mathcal{H}_+\to\mathcal{H}_+$ share one and the same set of
eigenvalues $\Lambda^2\in[0,1]$, and the orthonormal functions
 $\{\varphi_k\}$ and $\{\psi_k\}$
 constitute the orthonormal basis for
 the Hilbert spaces $\mathcal{H}_+$ and $\mathcal{H}_-$ respectively.
 The Hilbert-Schmidt norm of both operators,
\begin{equation}
\label{Hilbert_Schmidt}
\mathrm{tr}\left(P^\top_\mu P_\mu\right)=\mathrm{tr}\left(P_\mu P^\top_\mu \right)=
\sum_{s=1}^N\Lambda_s^2
\end{equation}
is a {\it global characteristic} of the directed graph.

Provided the random walks are defined on
 a strongly connected directed graph $\vec{G}$,
let us consider the functions $\rho^{(t)}(s)\in[0,1]\times\mathbb{Z}_+$ representing
the probability for finding a random walker at the node $s$, at time $t$.
A random walker located at the source node $s$ can reach the node
 $s'$ through either nodes.
 Being transformed in accordance to
(\ref{transf_pm}) and (\ref{transf_pm2}),  these
 function takes the following forms: $(\rho_s^{(t)})_{\mathcal{H}_-}=
 \mu_-^{-1/2}\rho^{(t)}(s)$
and
$(\rho_s^{(t)})_{\mathcal{H}_+}=\mu_+^{-1/2}\rho^{(t)}(s)$.
Then, the self-adjoint operators $U:\mathcal{H}_-\to\mathcal{H}_-$ and
$V:\mathcal{H}_+\to\mathcal{H}_+$ with
the matrix elements
\begin{equation}
\label{Matrix_elements}
\begin{array}{lcl}
U_{s\,s'}&=& \frac 1{\sqrt{k_\mathrm{out}(s)k_\mathrm{in}(s')}}
\sum_{i\in\vec{G}}
\frac{{A}^\top_{\vec{G}}{}_{is}{A}_{\vec{G}}{}_{is'}}{\sqrt{k_\mathrm{out}(i)
k_\mathrm{in}(i)}} ,\\
V_{s's}&=&
\frac 1{\sqrt{k_\mathrm{out}(s')k_\mathrm{in}(s)}}
\sum_{i\in\vec{G}}
\frac{{A}_{\vec{G}}{}_{s'i}{A}^\top_{\vec{G}}{}_{s\,i}}{\sqrt{k_\mathrm{out}(i)
k_\mathrm{in}(i)}}
\end{array}
\end{equation}
define the dynamical system
\begin{equation}
\label{dyn_system01}
\left\{
\begin{array}{lcl}
\sum_{s\in\vec{G}}\left(\rho_s^{(t)}\right)_{\mathcal{H}_-} U_{s\,s'} & =&
\left(\rho_{s'}^{(t+2)}\right)_{\mathcal{H}_-}, \\
\sum_{s'\in\vec{G}}\left(\rho_{s'}^{(t)}\right)_{\mathcal{H}_+} V_{s's}&
=&\left(\rho_{s}^{(t+2)}\right)_{\mathcal{H}_+}\quad .
\end{array}
\right.
\end{equation}

\subsection{Spectral analysis of self-adjoint operators defined on directed graphs}
\label{subsec:Spectral_analysis_direceted}
\noindent

 The spectral properties of the
self-adjoint operators $U$ and $V$ driving two Markov
 processes on strongly connected directed graphs and sharing
  the same non-negative eigenvalues $\Lambda^2_s\,\in\,[0,1]$
can be analyzed by the method of characteristic functions. Being
defined on a strongly connected directed graph the spectral
properties of self-adjoint operators $U$ and $V$ can be used in
order to investigate the structure of the graph precisely as it is
done in spectral graph
theory
for undirected graphs.

The main goal of morphological analysis being applied to directed
graphs is to detect the groups of nodes strongly correlated
with respect to the random traffic that arrives at and departs from
them. It is worth to mention that a node in a directed graph can
have dramatically different numbers of incoming and outgoing
links. As a consequence, nodes which could serve as a good source
of random traffic for many other nodes in the graph, at the same
time, could have an exponentially small probability to host a
random walker. The self-adjoint operators $U$ and $V$ describe
{\it correlations} between flows of random walkers entering and
leaving nodes in a directed graph. In the framework of spectral
approach,  they
 are labelled by
the eigenvalues $\Lambda^2_s$, $s=1,\ldots N$, and those correlations essential
for coherence of different segments of the transport network
correspond to the largest eigenvalues. The general method of PCA
(see Sec.~\ref{subsec:Graph_Partitioning}) can be applied
independently to $U$ and $V$ in order to detect segments of the
graph coherent with respect to
 $\mathcal{H}_+$ and $\mathcal{H}_-$.

The approach which we propose below
 for the analysis of coherent structures
that arise in strongly connected directed graphs is similar to
that one used in purpose of the spatiotemporal analysis of complex
signals  in \cite{Lima1991,Lima_Aubry93} and refers to the
Karhuen-Lo\'{e}ve decomposition in classical signal analysis
\cite{Karhunen}. In the framework of bi-orthogonal decomposition
discussed in \cite{Lima1991}, the complex spatio-temporal signal
has been decomposed into orthogonal temporal modes called {\it
chronos} and orthogonal spatial modes called {\it topos}. Then the
spectral analysis of the phase-space of the dynamics and the
spatio-temporal intermittency in particular has naturally led  to
the notions of "energies" and "entropies" (temporal, spatial, and
global) of signals. Each spatial mode has been associated with an
instantaneous coherent structure which has a temporal evolution
directly given by its corresponding temporal mode. In view of that
the thermodynamic-like quantities had been used in order to
describe the complicated spatio-temporal behavior of complex
systems.

In the present subsection, we demonstrate that a somewhat
similar approach can also be applied to the spectral analysis of
directed graphs. Namely, we will show
that the morphological structure of
directed graphs can be related to  quantities extracted from
bi-orthogonal decomposition of random walks defined on them.
Furthermore, in this context, the temporal modes and spatial
modes introduced in \cite{Lima1991} are the eigenfunctions of
correlation operators $U$ and $V$.  Although our approach can be
viewed as a version of the signal analysis, it is fundamentally
different from that in principle.
By definition
transition probability operators
satisfy the probability conservation property  but it is
in general  not the case for the spatio-temporal signals
generated by complex systems.

All coherent segments of a directed graph
 participate independently in the
Hilbert-Schmidt norm (\ref{Hilbert_Schmidt}) of
the self-adjoint operators $U$ and $V$,
\begin{equation}
\label{graph_energy}
\mathfrak{E}(\vec{G})=\sum_{s=1}^N\Lambda_s^2.
\end{equation}
Borrowing the terminology from theory of signals and
\cite{Lima1991}, we can call (\ref{graph_energy})  {\it energy},
the only additive characteristic of the directed graph $\vec{G}$.
While introducing the projection operators (in Dirac's notation)
by
\begin{equation}
\label{projectors}
\mathbb{P}^{(+)}_s=\left|\psi_s\rangle \langle\psi_s\right|,\quad
\mathbb{P}^{(-)}_s=\left|\varphi_s\rangle \langle\varphi_s\right|,\quad
s\,=\,1,\ldots N,
\end{equation}
we can decompose $\mathfrak{E}(\vec{G})$ into two components related to
the Hilbert spaces $\mathcal{H}_+$ and $\mathcal{H}_-$:
\begin{equation}
\label{energy_decomp}
\begin{array}{lcl}
\mathfrak{E}(\vec{G})&=&\mathfrak{E}^{(+)}(\vec{G})+\mathfrak{E}^{(-)}(\vec{G}) \\
 &=& \sum_{s=1}^N\,\Lambda_s^2\mathbb{P}^{(+)}_s\,+\,\sum_{s=1}^N\,\Lambda_s^2\mathbb{P}^{(-)}_s.
\end{array}
\end{equation}
Using (\ref{graph_energy}) as normalization, we can
consider the {\it relative energy} for each coherent structure of
the directed  graph by
\begin{equation}
\label{norm_energy}
\mathfrak{e}_s=\frac{\Lambda_s^2}{\mathfrak{E}(\vec{G})}
\end{equation}
and, in the spirit of \cite{Lima1991}, define the global entropy of coherent
structures in the graph $\vec{G}$ as
\begin{equation}
\label{graph_entropy}
\mathfrak{H}(\vec{G})=-\frac 1{\ln N}\sum_{s=1}^N \mathfrak{e}_s\ln \mathfrak{e}_s,\quad s=1,\ldots N,
\end{equation}
which is independent on the graph size $N $ due to the presence of
the normalizing factor $1/\ln N$, and can therefore be used in order to
 compare different directed graphs.
The global entropy of the graph $\vec{G}$ is zero if all its
 nodes belong to one and the same coherent structure
 (i.e., only one eigenvalue $\Lambda_s^2\ne 0$). In the opposite case,
$\mathfrak{H}(\vec{G})\to 1$ if most of eigenvalues $\Lambda_s^2$ are degenerate.

The relative energy (\ref{norm_energy}) can also be decomposed into
the $\mathcal{H}_+$- and $\mathcal{H}_-$-components:
\begin{equation}
\label{norm_energy_decomp}
\mathfrak{e}_s^{(\pm)}=\frac{\Lambda^2_s\mathbb{P}^{(\pm)}_s}
{\mathfrak{E}^{(\pm)}(\vec{G})},\quad s=1,\ldots N,
\end{equation}
and then the partial entropies can be defined as
\begin{equation}
\label{partial_entropies}
\mathfrak{H}^{(\pm)}(\vec{G})=-\frac 1{\ln N}\sum_{s=1}^N\,
\mathfrak{e}^{(\pm)}_s\ln \mathfrak{e^{(\pm)}}_s.
\end{equation}

The  conclusion is that any strongly connected
 directed graph $\vec{G}$ can be
considered as a bipartite graph with respect to the in- and
out-connectivity of nodes.
The bi-orthogonal decomposition of random walks is then used
 in order to define the self-adjoint operators on  directed
 graphs describing correlations between flows of random
  walkers which arrive at and leave the nodes.
  These self-adjoint operators share the non-negative real
   spectrum of eigenvalues, but different
    orthonormal sets of eigenvectors. The  standard
principal component analysis  can be applied also
 to directed graphs. The global characteristics
  of the directed graph and its components can
   be obtained from the  spectral properties
    of the self-adjoint operators.

\subsection{Self-adjoint operators for interacting networks}
\label{subsec:Interacting_netwroks}
\noindent

The  bi-orthogonal decomposition can also be implemented in order
to determine coherent segments of two or more interacting networks
defined on one and the same set of nodes $V$, $|V|=N$.

Given two different strongly connected weighted directed
graphs $\vec{G}_1$ and $\vec{G}_2$ specified on the
same set of $N$ vertices by the non-symmetric
adjacency matrices ${\bf A}^{(1)},$ ${\bf A}^{(2)}$, which entries
are the edge weights, $w^{(1,2)}_{ij}\geq 0$,
then the four transition operators of random walks can be defined on
both networks as
\begin{equation}
\label{4trasnitions}
{\bf P}^{(\alpha)} =
\left({\bf D}_\mathrm{out}^{(\alpha)}\right)^{-1}{\bf A}^{(\alpha)}, \quad
\left({\bf P}^{(\alpha)}\right)^{\star}
 = \left({\bf D}_\mathrm{in}^{(\alpha)}\right)^{-1}{\bf A}^{(\alpha)}, \quad \alpha=1,2,
\end{equation}
where ${\bf D}_{\mathrm{out/in}}$ are the diagonal matrices with the following entries:
\begin{equation}
\label{weights00deg}
k^{(\alpha)}_{\mathrm{out}}(j)= \sum_{i,{}j\to i} w^{(\alpha)}_{ji},\quad
k^{(\alpha)}_{\mathrm{in}}(j)= \sum_{i,{}i\to j} w^{(\alpha)}_{ij},\quad \alpha=1,2.
\end{equation}
We can define 4 different measures,
\begin{equation}
\label{4measures}
\begin{array}{ll}
\mu^{(1)}_-=\sum_j\,k^{(1)}_{\mathrm{out}}(j)\,\delta(j), &
\mu^{(1)}_+=\sum_j\,k^{(1)}_{\mathrm{in}}(j)\,\delta(j), \\
\mu^{(2)}_-=\sum_j\,k^{(2)}_{\mathrm{out}}(j)\,\delta(j),&
\mu^{(2)}_+=\sum_j\,k^{(2)}_{\mathrm{in}}(j)\,\delta(j).
\end{array}
\end{equation}
and four Hilbert spaces $\mathcal{H}^{(\alpha)}_{\pm}$
associated with the
spaces of square summable functions,
$\ell^2\left(\mu_{\pm}^{(\alpha)}\right)$, $ \alpha=1,2$.

Then the transitions operators
$P^{(\alpha)}_\mu:\mathcal{H}_-^{(\alpha)}\to
\mathcal{H}_-^{(\alpha)}$ and
$\left(P^{(\alpha)}_\mu\right)^\star:\mathcal{H}_+^{(\alpha)}\to
\mathcal{H}_+^{(\alpha)}$ adjoint with respect to the measures
$\mu_{\pm}^{(\alpha)}$ are defined by the following matrices:
\begin{equation}
\label{adjoint02}
\begin{array}{l}
\left(P^{(\alpha)}_\mu\right)_{ij}=\frac{A^{(\alpha)}_{\vec{G}}{}_{ij}}
{\sqrt{k^{(\alpha)}_\mathrm{out}(i)}\sqrt{k^{(\alpha)}_\mathrm{in}(j)}},
\\
\left(P^{(\alpha)}_\mu\right)^\star_{ij}=\frac{A^{(\alpha)}{}^\top_{\vec{G}}{}_{ij}}
{\sqrt{k^{(\alpha)}_\mathrm{in}(i)}\sqrt{k^{(\alpha)}_\mathrm{out}(j)}}.
\end{array}
\end{equation}
The spectral analysis of the above operators requires that four
equations be solved:
\begin{equation}
\label{two_eqs_02}
\left\{
\begin{array}{lcl}
{\bf P}^{(\alpha)}_\mu\varphi^{(\alpha)}&=&\Lambda^{(\alpha)}\psi^{(\alpha)},\\
{\bf P}_\mu^{(\alpha)}{}^\top\psi^{(\alpha)}&=&\Lambda^{(\alpha)}\varphi^{(\alpha)},
\end{array}
\right.
\end{equation}
where $\alpha=1,2$ as usual.

\begin{figure}[ht]
 \noindent
\begin{center}
\epsfig{file=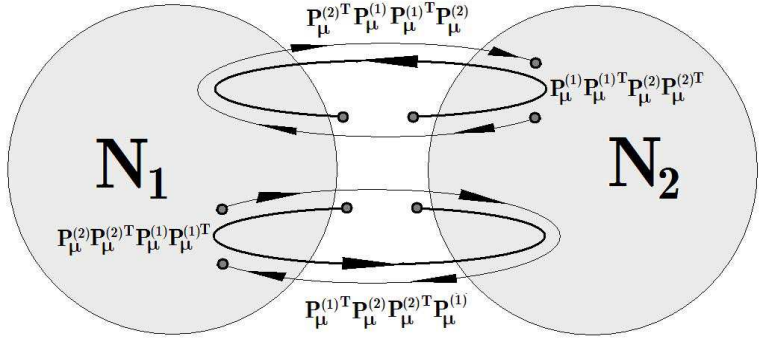, angle= 0,width =10cm, height =4.8cm}
  \end{center}
\caption{\small Self-adjoint operators for two interacting
networks sharing the same set of nodes. }
 \label{Fig2_19}
\end{figure}

 Any solution $\{\varphi^{(\alpha)},\psi^{(\alpha)}\}$ of
  the system (\ref{two_eqs_02}), up to the possible
partial isometries,
\begin{equation}
\label{isometries}
{\bf G}^{(\alpha)}\varphi^{(\alpha)}=\psi^{(\alpha)},
\end{equation}
also satisfies the system
\begin{equation}
\label{system_02}
\begin{array}{lcl}
{\bf P}^{(2)}_\mu{}^\top{\bf P}^{(1)}_\mu{\bf P}^{(1)}_\mu{}^\top{\bf P}^{(2)}_\mu\psi^{(2)}&=& \left(\Lambda^{(1)}\Lambda^{(2)}\right)^2 \psi^{(2)},\\
{\bf P}^{(1)}_\mu{}^\top{\bf P}^{(2)}_\mu{\bf P}^{(2)}_\mu{}^\top{\bf P}^{(1)}_\mu\psi^{(1)}&=& \left(\Lambda^{(1)}\Lambda^{(2)}\right)^2\psi^{(1)},\\
{\bf P}^{(1)}_\mu{\bf P}^{(1)}_\mu{}^\top{\bf P}^{(2)}_\mu{\bf P}^{(2)}_\mu{}^\top\varphi^{(1)}&=& \left(\Lambda^{(1)}\Lambda^{(2)}\right)^2\varphi^{(1)},\\
{\bf P}^{(2)}_\mu{\bf P}^{(2)}_\mu{}^\top{\bf P}^{(1)}_\mu{\bf P}^{(1)}_\mu{}^\top\varphi^{(2)}&=& \left(\Lambda^{(1)}\Lambda^{(2)}\right)^2\varphi^{(2)},\\
\end{array}
\end{equation}
in which $\left(\Lambda^{(1)}\Lambda^{(2)}\right)^2\in[0,1].$
Operators in the l.h.s of the system (\ref{system_02}) describe
correlations between flows of random walkers which go through
vertices following the links of either networks. They are
represented by the non-negative symmetric matrices with
orthonormal eigenvectors which can be subjected to the standard
methods of the PCA analysis. Their spectrum
 can also  be investigated by the methods
discussed in the previous subsection.

It is convenient to represent the self-adjoint operators from
the l.h.s. of (\ref{system_02}) by the closed directed
 paths shown in the diagram in Fig.~\ref{Fig2_19}.
Being in the self-adjoint products of transition
operators, $P_\mu^{(\alpha)}$ corresponds to the flows
 of random walkers which depart from either networks,
and  $P_\mu^{(\alpha)}{}^\top$ is for those which arrive at the
network $\alpha$. From Fig.~\ref{Fig2_19} , it is clear that the
self-adjoint operators in (\ref{system_02}) represent all possible
closed trajectories visiting both networks $\mathbf{N}_1$ and
${\bf N}_2$.

In general, given a complex system consisting
 of $n>1$ interacting networks operating on the
same set of nodes, we can define $2^n$ self-adjoint operators
related to the different modes of random walks. Then the set of
network nodes can be separated into a number of essentially
correlated
 segments
with respect to each of self-adjoint operators.

\section{Discussion and Conclusion}
\label{sec:Discussion}
\noindent

In the present paper, we have developed a self-consistent
approach to complex transport networks based on the use of Markov
chains defined on their graph representations.

From Euler's time, urban design
and townscape studies
were the sources of inspiration
for the network analysis and graph theory.
It is common now that
networks are
the reality of urban renewals \cite{renewal}.
Flows of pedestrians
and vehicles
through a city  are
dependent on one another
and
that
requests for organizing
them in a network setting.

The networking is structurally contagious.
In order to
be able to
 master a network
effectively,
an authority
 should also constitute
a network structure,
probably
as complicated as the one that
it supervises.
The
governance and maintenance
units supporting urban
 network renewals,
negotiating performance targets,
taking decisions,
financing and eventually
implementing them also
form  networks.
A complex network of city itineraries
that we can experience in everyday life
appears as
the result of multiple
complex interactions
between a number of transport,
 social,
and economical networks.

It is intuitively clear that
a complex network
in equilibrium
emerges from synergy and interplay
 between
the {\it topological structure}
 shaped by
a connected graph $G$
with some
 positive
{\it measures} (masses)
 appointed for the vertices
and some
 positive {\it weights}
assigned to the edges,
the {\it dynamics}
described by
the set of operators defined on $G$,
and the properties of
the {\it embedding physical space}
 specified by
the metric length distances of the edges.

The approach we have discussed in the present paper helps to define an
equilibrium state for complex transport networks and investigate its
properties.

\section{Acknowledgment}
\label{Acknowledgment}
\noindent

The work has been supported by the Volkswagen Foundation (Germany)
in the framework of the project: "Network formation rules, random
set graphs and generalized epidemic processes" (Contract no Az.:
I/82 418). The authors acknowledge the multiple fruitful
discussions with the participants of the workshop {\it Madeira
Math Encounters XXXIII}, August 2007, CCM - CENTRO DE CI\^{E}NCIAS
MATEM\'{A}TICAS, Funchal, Madeira (Portugal).


\begin{thebibliography}{000}



\bibitem{Braess}
D. Braess, "\"{U}ber ein Paradoxon aus der Verkehrsplannung".
{\it Unternehmensforschung} {\bf 12}, 258-268 (1968).
\bibitem{Kolata}
G. Kolata, "What if they closed the $42^{\mathrm{nd}}$ Street and nobody noticed?", {\it  The New
York Times}, Dec. 25  (1990).

\bibitem{Manning}
A. Manning, {\it Ann. of Math.} {\bf 110}, 567-573 (1979).
\bibitem{Bourdon}
M. Bourdon, {\it L'Einseign. Math.,} {\bf 41} (2), 63-102 (1995) (in French).
\bibitem{Roblin}
T. Roblin, {\it Ann. Inst. Fourier} (Grenoble) {\bf 52}, 145-151 (2002) (in French).


\bibitem{Lim:2005}
S. Lim,  {\it Minimal Volume Entropy on Graphs}. Preprint arXiv:math.GR/050621,
 (2005).

\bibitem{Wardrop:1952}
 J.G. Wardrop, {\it Proc. of the Institution of Civil
Engineers} {\bf 1} (2), pp. 325-362 (1952).

\bibitem{Beckmann:1956}
M.J. Beckmann, C.B. McGuire, C.B.  Winsten,
{\it Studies in the Economics of Transportation.}
Yale University Press, New Haven, Connecticut (1956).

\bibitem{Kuipers}
 B. Kuipers, {\it Environment and Behavior}   {\bf 14} (2),
  pp. 202-220 (1982).

\bibitem{Hillier:1984}
B. Hillier,   J. Hanson,  {\it The Social Logic of Space}. Cambridge University Press. ISBN 0-521-36784-0 (1984).
 \bibitem{Hillier:1999}
B. Hillier, {\it Space is the Machine: A Configurational Theory of Architecture}. Cambridge University Press. ISBN 0-521-64528-X (1999).
\bibitem{Penn:2001}
 A.  Penn, Space Syntax and Spatial Cognition. Or, why the axial line? In: Peponis, J. and Wineman, J. and Bafna, S., (eds). {\it Proc. of the Space Syntax $3^{rd}$ International Symposium}, Georgia Institute of Technology, Atlanta, May 7-11 2001.

\bibitem{Smola2003}
A. Smola and R. I. Kondor. "Kernels and regularization on graphs". In {\it Learning
    Theory and Kernel Machines}, Springer (2003).

\bibitem{Aldous}
D.J. Aldous, J.A. Fill,  {\it Reversible Markov Chains and Random Walks on Graphs}.
A book in preparation, available at www.stat.berkeley.edu/aldous/book.html.

\bibitem{Lovasz:1993}
 L. Lov\'{a}sz, {\it Bolyai Society Mathematical
 Studies} {\bf 2}: {\it Combinatorics,
Paul Erd\"{o}s is Eighty}, Keszthely (Hungary), p. 1-46 (1993).

\bibitem{Chung:1997}
F. Chung, {\it Lecture notes on spectral graph theory}, AMS Publications Providence (1997).

\bibitem{Vision}
T. Morris, {\it Computer Vision and Image Processing}. Palgrave Macmillan.
 ISBN 0-333-99451-5 (2004).

\bibitem{PCA}
K. Fukunaga, {\it Introduction to Statistical Pattern Recognition}, ISBN 0122698517,
Elsevier (1990).

\bibitem{PCA2}
I.T. Jolliffe, {\it Principal Component Analysis} (2-nd edition) Springer Series in
Statistics (2002).
\bibitem{Pearson}
 J. Cohen,  P. Cohen,  S.G. West, L.S. Aiken,
  {\it Applied multiple regression/correlation analysis for the behavioral sciences}.
  (3rd ed.) Hillsdale, NJ: Lawrence Erlbaum Associates (2003).
\bibitem{Lafon}
B. Nadler, S. Lafon, R.R. Coifman and I.G. Kevrekidis, "Diffusion
maps, spectral clustering and reaction coordinate of dynamical
systems",
 {\it Applied and Computational Harmonic Analysis: Special
 issue on Diffusion Maps and Wavelets}, Vol. {\bf 21},113-127 (2006).

\bibitem{Kondor_Laff}
R. I. Kondor and J. Lafferty, "Diffusion kernels on graphs and other discrete
structures". In C. Sammut and A. G. Hoffmann, editors, {\it Machine Learning},
Proceedings of the 19th International Conference (ICML 2002), pages 315-322.
San Francisco, Morgan Kaufmann, (2002).

\bibitem{Zhu}
X. Zhu, Z. Ghahramani, J. Lafferty, "Semi-supervised learning using gaussian fields and
harmonic functions". In Proc. {\it 20th International Conf. Machine Learning}, vol.
{\bf 20}, 912 (2003).
\bibitem{Saitoh}
S. Saitoh, {\it Theory of Reproducing Kernels and its Applications}, Longman
Scientific and Technical, Harlow, UK (1988).

\bibitem{Wahba}
G. Wahba, {\it Spline Models for Observational Data}, Vol. {\bf 59} of CBMS-NSF
{\it Regional Conference Series in Applied Mathematics}, SIAM, Philadelphia (1990).
\bibitem{StatMech}
Shang-Keng Ma, {\it Statistical mechanics}, World Scientific (1985).

\bibitem{Volchenkov:2007}
D. Volchenkov,  Ph. Blanchard, {\it Physical Review E} {\bf 75}(2), id 026104 (2007).

\bibitem{AlbertBarabasi2002}
R. Albert, A.L. Barab\'{a}si, {\it Rev. Mod. Phys.} {\bf 74}, 47 (2002).
\bibitem{Bianconi}
G. Bianconi, N. Gulbahce, A.E. Motter, {\it Local structure of directed networks}, E-print arXiv:0707.4084 [cond-mat.dis-nn] (27 July 2007).
\bibitem{WattsStrogatz}
D.J. Watts, S.H. Strogatz, {\it Nature} {\bf 393}(6684), 440-442 (1998).

\bibitem{Chung2005}
F. Chung, {\it Annals of Combinatorics} {\bf 9}, 1-19 (2005).
\bibitem{LovWinkl}
L. Lov\'{a}sz, P. Winkler, {\it Mixing of Random Walks and Other Diffusions on a Graph}.
Surveys in combinatorics, Stirling, pp. 119–154 (1995); London Math. Soc. Lecture
Note Ser., vol. {\bf 218}, Cambridge Univ. Press.

\bibitem{chip-firing}
A. Bj\"{o}ner, L. Lov\'{a}sz, P. Shor, {\it Europ. J. Comb.} {\bf 12}, 283-291 (1991).%

\bibitem{chip-firing2}
A. Bj\"{o}ner, L. Lov\'{a}sz, {\it J. Algebraic Comb.} {\bf 1}, 305-328 (1992).
\bibitem{Butler2}
S. Butler, {\it Electronic Journal of Linear Algebra}, {\bf 16} 90 (2007).

\bibitem{Lima1991}
N. Aubry, R. Guyonnet, R. Lima, {\it J. Stat. Phys.} {\bf 64}, 683-739 (1991).
\bibitem{Aubry1991}
N. Aubry, {\it Theor. and Comp. Fluid Dyn.} {\bf 2}, 339-352 (1991).


\bibitem{Karhunen}
K. Karhunen, {\it Ann. Acad. Sci. Fennicae} {\bf A:1} (1944).
\bibitem{Loeve}
M. Lo\`{e}ve, {\it Probability Theory}, van Nostrand, New York (1955).

\bibitem{Lima_Aubry93}
N. Aubry, L. Lima, {\it Spatio-temporal symmetries}, Preprint CPT-93/P.2923, Centre de Physique Theorique, Luminy, Marseille, France (1993).
\bibitem{renewal}
M. Haffner, M. Elsinga, {\it
Urban renewal performance in complex networks
Case studies in Amsterdam North and Rotterdam South},
W16 – Institutional and Organizational Change in Social Housing
Organization in Europe, Int. Conference on Sustainable Urban Areas, Rotterdam (2007).


\end{thebibliography}
\end{document}